\documentclass[letterpaper,journal]{IEEEtran}
\usepackage{amsmath,amsfonts}
\usepackage{algorithmic}
\usepackage{algorithm}
\usepackage{array}
\usepackage[caption=false,font=normalsize,labelfont=sf,textfont=sf]{subfig}
\usepackage{textcomp}
\usepackage{stfloats}
\usepackage{url}
\usepackage{verbatim}
\usepackage{graphicx}
\usepackage{cite}
\usepackage{multirow}

\newcommand*{\defeq}{\mathrel{\vcenter{\baselineskip0.5ex \lineskiplimit0pt
                     \hbox{\scriptsize.}\hbox{\scriptsize.}}}%
                     =}

\DeclareMathOperator*{\argmax}{arg\,max}

\begin{document}

\title{Privacy-Aware Load Balancing in Fog Networks:\\ A Reinforcement Learning Approach}

\author{Maad Ebrahim, Abdelhakim Hafid,~\IEEEmembership{Member,~IEEE}% <-this % stops a space
\thanks{The authors are with the NRL, Department of Computer Science and Operational Research, University of Montreal, Montreal, QC H3T-1J4, Canada (e-mail: maad.ebrahim@umontreal.ca; ahafid@iro.umontreal.ca).}% <-this % stops a space
\thanks{Corresponding author: Maad Ebrahim (maad.ebrahim@umontreal.ca).}}

\maketitle

\begin{abstract}
% Fog Computing has emerged as a solution to support the growing demands of real-time Internet of Things (IoT) applications, which require high availability of such distributed services. Intelligent workload distribution algorithms are needed to maximize the utilization of Fog resources while minimizing the time required to process these workloads. These load balancing algorithms are critical in dynamic environments with heterogeneous resources and workload requirements along with unpredictable traffic demands. In this paper, load balancing is provided using a Reinforcement Learning (RL) algorithm, which optimizes the system performance by minimizing the waiting delay of IoT workloads. Unlike previous studies, the proposed solution does not require load and resource information from Fog nodes, which makes the algorithm dynamically adaptable to possible environment changes over time. This also makes the algorithm aware of the privacy requirements of Fog service providers, who might like to hide such information to prevent competing providers from calculating better pricing strategies. The proposed algorithm is interactively evaluated on a Discrete-event Simulator (DES) to mimic a practical deployment of the solution in real environments. In addition, we evaluate the algorithm's generalization ability on simulations longer than what it was trained on, which, to the best of our knowledge, has never been explored before. The results provided in this paper show how our proposed approach outperforms baseline load balancing methods under different workload generation rates.
In this paper, we propose a load balancing algorithm based on Reinforcement Learning (RL) to optimize the performance of Fog Computing for real-time IoT applications. The algorithm aims to minimize the waiting delay of IoT workloads in dynamic environments with unpredictable traffic demands, using intelligent workload distribution. Unlike previous studies, our solution does not require load and resource information from Fog nodes to preserve the privacy of service providers, who may wish to hide such information to prevent competitors from calculating better pricing strategies. The proposed algorithm is evaluated on a Discrete-event Simulator (DES) to mimic practical deployment in real environments, and its generalization ability is tested on simulations longer than what it was trained on. Our results show that our proposed approach outperforms baseline load balancing methods under different workload generation rates, while ensuring the privacy of Fog service providers. Furthermore, the environment representation we proposed for the RL agent demonstrates better performance compared to the commonly used representations for RL solutions in the literature, which compromise privacy. 
\end{abstract}

\begin{IEEEkeywords}
Internet of Things, Cloud Computing, Fog Computing, Edge Computing, Task Assignment, Service Selection, Load Balancing, Reinforcement Learning.
\end{IEEEkeywords}

\section{Introduction}
\label{sec:intro}

Fog Computing is a technology that offers distributed computing resources very close to IoT devices, using resource-rich routers for example. It fits IoT applications better than Cloud Computing as it supports delay-sensitive, geo-distributed, location-aware, and mobile applications \cite{Survey2015}. In addition, it reduces the traffic between end devices and the Cloud, and hence, increases the security and privacy of IoT applications by pre-processing data closer to its source \cite{SecurityPrivacy}. These IoT applications must be modular to benefit from the distributed nature of Fog Computing, i.e., pipelined workflows \cite{Vertical}.

Resource management is critical to cope with the increasing demands of end devices and the scarce resources in Fog nodes. In addition, fluctuations in data generation rates and mobility can dynamically cause uneven distribution of IoT devices (e.g., sensors) in relation to Fog resources. Hence, distributing the load optimally between Fog resources is essential to improve the system performance, i.e., increasing resource utilization and reducing the response time. This can be done by avoiding bottlenecks, overloads, and underloads \cite{SystematicReview}. To do this, the amount of available Fog resources should always be greater than the requirements of the workloads being generated \cite{LBEdge}. Besides improving the system performance, resource management solutions should maintain the privacy requirements of trustless multi-provider Fog systems to avoid the need for complex privacy solutions \cite{Blockchain}.

Reinforcement Learning (RL) can tackle control problems in dynamic partially-observable environments \cite{sutton2018reinforcement}. Balancing the load in Fog environments is an example of such control problems, where an RL agent continually adapts to environment changes by learning from its own experience in the environment. RL agents do not necessarily need to know the system model (transition dynamics), and can achieve optimal, or near optimal, solutions without the need for private, complex, or unknown information from the environment. Deep RL (DRL) \cite{mnih2013playing} incorporate function approximation of Deep Neural Networks (DNN) into RL algorithms to solve problems with large state spaces. This is achieved by the generalization ability of DNN to unseen states, where almost every state encountered will never have been seen before. RL models require considerable amount of time and resources to be trained, but then, light versions of the trained models can be extracted and run with minimal compute and storage requirements.

Therefore, we propose in this paper a privacy-aware load balancing (LB) solution for Fog networks using Double Deep Q-Learning (DDQL). DDQL is a DRL algorithm that uses two separate neural networks: one to select the action and another to evaluate it. In most cases, Fog service providers prefer not to share load and resource information from their Fog nodes; this information can be used by competing service providers to determine competing pricing strategies. However, existing solutions often require Fog resource and/or load information (e.g., \cite{ReTra, EnergyEfficient, smartcities, LBOS, Managing, HeterogeneousTask}), which requires the agent to retrain in case of dynamic changes in the environment. In this paper, our DDQL agent does not require load and resource information to optimally distribute the load between Fog nodes. Instead, it only needs to know the number of waiting jobs in the queues of those nodes without requiring any other information, e.g., computation and storage capacity. The key contribution of this paper can be summarized as follows:
\begin{itemize}
\item Our proposed DDQL algorithm provides optimal load distribution by minimizing the total number of waiting jobs in the queues of Fog nodes.
\item It provides privacy for Fog service providers as load and resource information of Fog nodes are not required. Instead, it works by observing the change in the number of jobs currently waiting in Fog nodes as an immediate reward for the agent's last action.
\item By not requiring resource and load information, the proposed algorithm adapts to dynamic changes in the environment. In addition, it adapts to dynamic changes in workload requirements and the distribution of IoT devices. This is because it works with workload source clusters and discrete categories of workload requirements.
\item The proposed solution was interactively evaluated using a Discrete-event Simulator (DES), which mimics practical deployment in real environments. Unlike existing solutions, where the system model is hard-coded inside the agent step function.
\item To demonstrate actual deployment in real environments, our agent is evaluated with more simulation steps than what is trained with. In practical deployment, the agent is only retrained in case of new changes in the environment.
\end{itemize}

The rest of the paper is organized as follows. Section \ref{sec:related} presents related work. Section \ref{sec:rl} presents an overview of the RL algorithm. Section \ref{sec:method} presents the proposed DDQL LB solution. Section \ref{sec:evaluation} evaluates the performance of our approach compared to common baselines. Finally, Section \ref{sec:conclusion} concludes the paper.

\section{Related Work}
\label{sec:related}

Learning by interaction made RL the best approach to solving control problems in dynamic systems \cite{sutton2018reinforcement}, including LB problems \cite{NetworkDRLSurvey}. Load distribution is critical for complex computational networks like Fog Computing systems. Optimal load distribution in Fog systems must improve the overall system performance, i.e., minimizing request processing time. Minimizing the processing delay leads to fewer idle and fewer overclocked nodes, which results in an increase in resource utilization and a decrease in energy consumption.

RL has been used to minimize the average execution delay in Fog environments \cite{RL_RM_Fog}. However, numerical experimentations have been often used to evaluate these approaches, which do not reflect practical deployment. Divya et al. \cite{ReTra} evaluated their Q-Learning LB solution on a testbed with high volume simulated data. They were able to balance the load in isolated Fog clusters (each cluster is assigned to a set of end devices) wasting resources, and hence energy, in other clusters. This happens when there is a significant difference in the demand, in each cluster, relative to their corresponding resources.

Swarup et al. \cite{EnergyEfficient} used a DES simulator to interactively evaluate their DDQL LB solution. DES evaluation is more practical as it helps with result reproduction, benchmarking, and algorithm improvements. However, they considered isolated Fog nodes that are only interconnected through the Cloud; each Fog node is connected to a number of Edge nodes and each Edge node is connected to a number of end devices. Their goal was to balance the load in the Virtual Machines (VMs) of each node, using their CPU, RAM, and Storage information, instead of balancing the load between the nodes themselves.

AlOrbani et al. \cite{smartcities} proposed a Q-Learning agent that balances the load based on the internal characteristics of each VM, i.e., task communication, waiting, and processing delays. They also added a parameter that represents resource availability in each VM. This approach is not applicable to situations where service providers do not allow publicly sharing load and resource information about their Fog nodes. In addition, requiring Fog resource information does not allow for dynamic adaptation to dynamically changing environments. Indeed, the agent will need to detect these changes, i.e., resource upgrade or downgrade, and retrain every time a change is detected.

Similarly, Talaat et al. \cite{LBOS} used Q-Learning with a genetic algorithm to allocate and reallocate (migrate) jobs based on the load in Fog nodes, i.e., cache, RAM, and CPU usage. They also used a genetic algorithm to optimize an adaptive weighted metric based on those three features. They used a three-layer Fog architecture with a master controller for every Fog region; each controller runs its own agent to balance the load in its region. They evaluated their approach using numerical simulations, similar to \cite{smartcities, Managing, HeterogeneousTask}; however, they use a public healthcare dataset to generate the data.

Baek et al. \cite{Managing} used Q-Learning to minimize processing delay and overload probability by offloading the tasks between Fog nodes. For each Fog node, they find the optimal number of tasks to offload, and the optimal Fog node to be selected. They used a realistic flat Fog architecture, where Fog nodes can directly intercommunicate without the need to pass through the Cloud. The agent gets a reward for every offloading decision based on communication, waiting, and processing delays, as well as the maximum queue capacity of Fog nodes.

Baek et al. \cite{HeterogeneousTask} used the same Fog architecture (5 interconnected Fog nodes) and the same reward function (using resource and load information) as in \cite{Managing}. However, in \cite{HeterogeneousTask}, they used Deep Recurrent Q-Learning to decide between processing tasks locally, offloading them to a neighboring Fog node, or offloading them to the Cloud. Their agent takes these decisions based on predefined load categories, i.e., delay-critical, delay-sensitive, and delay-tolerant, which simulate load heterogeneity.

Under the scope of the above challenges, there is a need to consider the privacy of Fog nodes in terms of their resource and load information, which helps the LB algorithm to adapt automatically to environmental changes. In addition, realistic evaluation of LB algorithms must be done interactively on practical architectures instead of numerical experimentations. To fill out the gaps in the literature, our proposed approach takes into consideration the privacy of both load and resource information of Fog nodes. Not using Fog resource information by our agent makes it resilient to possible changes in these resources. Furthermore, we evaluate our approach on a realistic flat Fog architecture with randomly interconnected nodes; it is implemented using a DES simulator to mimic a realistic deployment scenario. In addition, we evaluate the generalization ability of our agent on longer episodes, relative to what it was trained on, which has never been considered before.

% \begin{table*}[!t]
% \scriptsize
% \caption{Comparison between existing RL-based LB approaches to optimize Fog environments \label{tab:compare}}
% \centering
% \begin{tabular}{|l|c|c|c|c|c|}
% \hline
% \textbf{Study} & \textbf{Privacy-Aware} & \textbf{Dynamic Adaptation} & \textbf{Realistic Architecture} & \textbf{Interactive Evaluation} & \textbf{Generalization Evaluation}\\
% \hline
% Baek et al. \cite{Managing} & & & \checkmark & & \\
% \hline
% Baek et al. \cite{HeterogeneousTask} & & & \checkmark & & \\
% \hline
% AlOrbani et al. \cite{smartcities} & & & & & \\
% \hline
% Divya et al. \cite{ReTra} & & & \checkmark & \checkmark & \\
% \hline
% Talaat et al. \cite{LBOS} & & & & & \\
% \hline
% Swarup et al. \cite{EnergyEfficient} & & & & \checkmark & \\
% \hline
% \textbf{Our Proposed Approach} & \checkmark & \checkmark & \checkmark & \checkmark & \checkmark \\
% \hline
% \end{tabular}
% \end{table*}

\section{The RL Algorithm}
\label{sec:rl}
The LB problem can be formulated as a Markov decision process (MDP). For every generated workload, an agent observes the current state of the system ($s \in S$) to take an action ($a \in A$), which forms a transition to a new state $s'$ with an immediate reward $r$. Equation \ref{eq:p} represents the probability of moving to the new state ($s'$) from state ($s$) after taking action ($a$). The reward generated by this transition represents an evaluation of that action. The accumulative performance of the behavior of the agent can be then represented through the expected return ($G$) of a series of actions (see Equation \ref{eq:g}). $\gamma$ defines the importance of future rewards while calculating the expected return. This expectation can be used to evaluate the expected value of future states after following the current policy of the agent.
\begin{equation}\label{eq:p}
    P_a(s, s') = Pr(s_{t+1}=s' \mid s_t=s, a_t=a)
\end{equation}
\begin{equation}\label{eq:g}
    G = \sum_{t=0}^{\infty} \gamma^t r_t \enspace , \qquad\gamma\in\left[0, 1\right]
\end{equation}

A stochastic policy ($\pi$) represents the probability of taking a given action ($a$) at a given state ($s$) (see Equation \ref{eq:a}). RL agents try to learn the optimal policy ($\pi^*$), which maximizes the total expected discounted reward, i.e., value function $V_\pi(s)$. The value function estimates how good it is to be in a given state and then following the current policy of the agent (see Equation \ref{eq:v}). It can be thought of as a prediction of how much reward the agent will get from the next state if it follows its current policy. Hence, the value function using the optimal policy, i.e., optimal value function $V^*(s)$, cannot be less than the value function using any other policy (see Equation \ref{eq:vpi}).
\begin{equation}\label{eq:a}
    \pi(a, s) = Pr(a_t=a \mid s_t=s)
\end{equation}
\begin{equation}\label{eq:v}
    V_\pi(s) = \mathbb{E}\left[G \mid s, \pi \right]
\end{equation}
\begin{equation}\label{eq:vpi}
    V^*(s) = \max_\pi V_\pi(s) = V_{\pi^*}(s) \geq V_\pi(s)
\end{equation}

Given a state ($s$), an action ($a$), and a policy ($\pi$), we can evaluate how good it is to take action ($a$) in state ($s$) and follow policy ($\pi$) thereafter. This is called the Q-Value, Q-Function, or action-value function, and it represents the total expected discounted reward for each state-action pair (see Equation \ref{eq:q}). When referring to finding optimal policies, action-value functions are tightly related to the state-value functions via Bellman optimality equations. In fact, the state-value function can be obtained from the $Q$-Function by selecting the action that maximizes the $Q$-Value for each state (see Equation \ref{eq:qv}).
\begin{equation}\label{eq:q}
    Q_\pi(s, a) = \mathbb{E}\left[G \mid s, a, \pi \right]
\end{equation}
\begin{equation}\label{eq:qv}
    V(s) = \max_aQ(s, a) \qquad \forall s \in S 
\end{equation}

The LB problem is an infinite-horizon problem, where there is no final decision step, i.e., terminal state. These types of problems require model-free algorithms, which are also useful with unknown/complex reward functions and transition dynamics. Temporal Difference (TD) Learning methods and Function approximation can be used to solve infinite-horizon MDPs \cite{sutton2018reinforcement}. With function approximation (Equation \ref{eq:qa}), a mapping of state-action pairs ($\phi$) is weighted by $\theta$ to calculate the action-values of each state-action pair. TD methods can be also used in continuous environments with no episodes as they use incomplete returns to update the values of each state. TD(0), for example, uses one-step look-ahead updates while the updates of TD(n) algorithms depend on n-step look-aheads.
\begin{equation}\label{eq:qa}
    Q(s, a) = \sum_{i=1}^d \theta_i \phi_i (s, a)
\end{equation}

Q-Learning \cite{watkins1992q} is one of the most common TD-based model-free RL algorithms that works with continuous states and discrete actions. Equation \ref{eq:ql} shows how Q-Learning is a TD-based algorithm, where the new $Q$-Value is calculated based on the old $Q$-Value and a TD error that is multiplied by the learning rate ($\alpha$). The TD error is computed by subtracting the old $Q$-Value from a temporal difference target which is calculated by adding the immediate reward to the discounted estimation of the optimal future value.
\begin{equation}\label{eq:ql}
    Q(s, a) \leftarrow Q(s, a) + \alpha \overbrace{\left[ \underbrace{r + \gamma \overbrace{\max_{a'}Q(s', a')}^{\text{Estimation of future value}}}_{\text{Temporal difference target}} - \enspace Q(s, a) \right]}^{\text{TD error}}
\end{equation}

The $Q$-Function is used to find the best action in a given state by searching for the action that maximizes the $Q$-Value for that state (see Equation \ref{eq:qpa}). Likewise, the agent experience, i.e., learned policy, is defined for every state, i.e. $\pi(s)$, as shown in Equation \ref{eq:qp}. Once the algorithm converges to the optimal $Q$-Value, i.e., $Q^*$, the optimal policy ($\pi^*$) can be then obtained similarly.
\begin{equation}\label{eq:qpa}
    \pi(a \mid s) = \argmax_aQ(s, a)
\end{equation}
\begin{equation}\label{eq:qp}
    \pi(s) = \argmax_aQ(s, a) \qquad \forall s \in S 
\end{equation}

To avoid implementing and searching through huge $Q$ tables in Q-Learning, DRL uses DNNs to approximate the Action-Value ($Q$), i.e., $Q$-Network. This helps for better generalization in infinite-horizon MDPs with continuous state-spaces, as in Deep Q-Learning (DQL). However, updating the $Q$-Network by a single step might significantly oscillate the policy of the agent, causing instability and divergence. To mitigate this, a technique called Experience Replay is introduced, where actions from a random history of samples are used to update the $Q$-Network instead of the most recent step \cite{mnih2013playing}.

Double-DQL (DDQL) was introduced to solve over-estimating the Action-Value ($Q$), a problem in Q-Learning and DQL algorithms \cite{ddqn}. It uses a model $Q$ and a target model $Q'$, where the parameters of $Q$ are periodically copied to $Q'$ every predefined number of steps. The agent uses $Q'$ for action selection while $Q$, which is updated on every training step, is only used to evaluate the actions making the algorithm off-policy. Being off-policy helps balance exploration and exploitation by learning the optimal policy while following an exploratory policy to choose the best action.

\section{Methodology}
\label{sec:method}

A set of $N$ nodes are used in our system; each node defined by its compute ($IPT_x$) and memory ($RAM_x$) resources (see Equation \ref{eq:N}). Nodes are connected through $L$ bidirectional links; each link is characterized by the pair ($n_i$, $n_j$) that it connects, its bandwidth $BW_x$, and its propagation delay $PR_x$ (see Equation \ref{eq:L}). Equation \ref{eq:App} shows the set of distributed applications, i.e., distributed data flow (DDF) models \cite{7356560}, that simultaneously run in the system; each application is represented by a set of modules and a set of dependencies between these modules. Our Fog environment is then defined in Equation \ref{eq:E} by the set of nodes, links, and distributed applications running in the system.
\begin{equation}\label{eq:N}
    N = n_1, n_2, \cdots, n_z, \quad \text{where} \enspace n_x \defeq \langle IPT_x, RAM_x \rangle
\end{equation}
\begin{equation}\label{eq:L}
    L = l_1, l_2, \cdots, l_z , \quad \text{where} \enspace l_x \defeq \langle n_i, n_j, BW_x, PR_x \rangle
\end{equation}
\begin{equation}\label{eq:App}
    DA = da_1, da_2, \cdots, da_z , \quad \text{where} \enspace da_x \defeq \langle M_x, D_x \rangle
\end{equation}
\begin{equation}\label{eq:E}
    E \defeq \langle N, L, DA \rangle
\end{equation}

\begin{figure}[!t]
\centering
\includegraphics[trim=0 200 0 200, clip, width=0.48\textwidth]{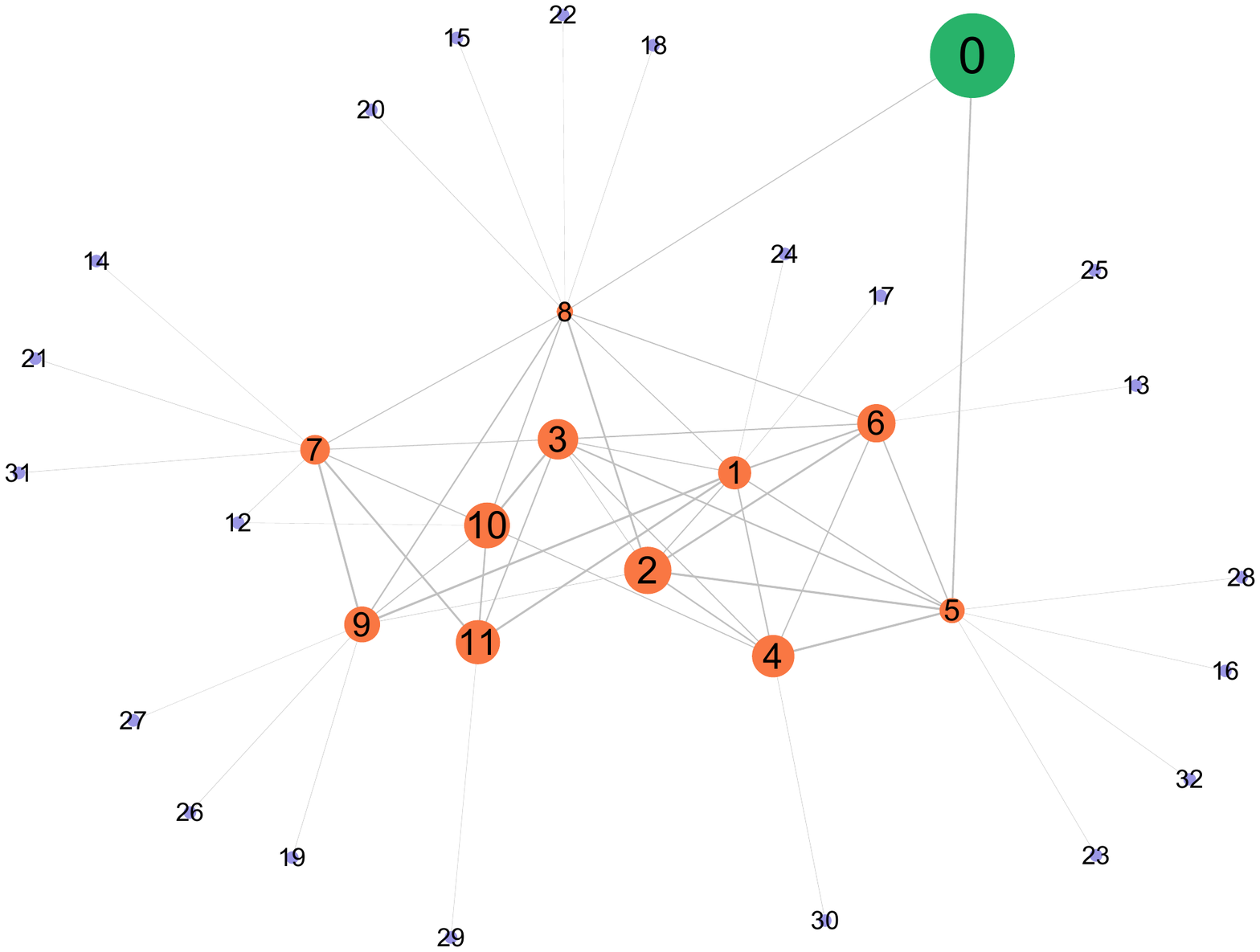}
\caption{Unbalanced Fog environment with IoT clusters (purple), Fog nodes (orange), and a Cloud (green) \cite{ELECTRE}.}
\label{fig:topo2}
\end{figure}

Realistic Fog topologies with unbalanced Fog resources and unbalanced workload distribution were simulated using a DES simulator (see Fig. \ref{fig:topo2}). Heterogeneous resources and workload requirements were considered in a non-hierarchical architecture. Since large number of IoT devices are often geographically deployed together, we can simulate them as a single cluster of IoT devices connected to the system through a single link (see Fig. \ref{fig:topo2}). Distributed applications with different compute requirements simultaneously run in our system, each with two flows of messages, i.e., Fog and Cloud feedback. In realistic applications, immediate Fog feedback is often needed for every source workload. The Cloud is only involved to perform data aggregation, with the possibility to provide feedback based on the aggregated data. The reader can refer to our previous work \cite{ELECTRE} for more details about the evaluation environment.

\begin{figure}[!t]
\centering
\includegraphics[trim=130 80 130 120, clip, width=0.48\textwidth]{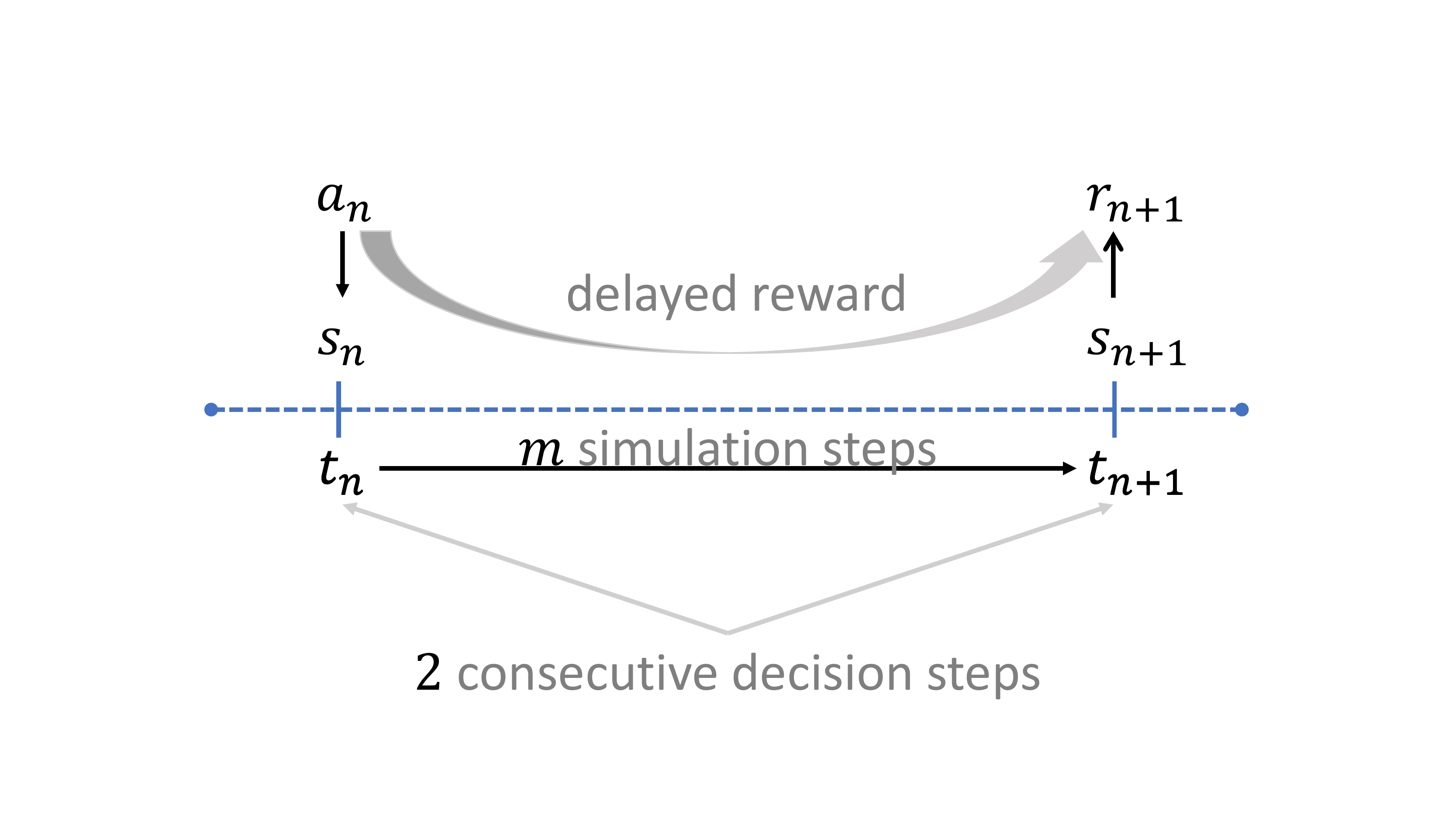}
\caption{Simulation steps vs. decision time-steps.}
\label{fig:reward}
\end{figure}

The proposed DDQL agent was implemented using TF-Agents \cite{TFAgents} to interactively work with the DES environment. First, we need to differentiate between simulation steps and decision steps (see Fig. \ref{fig:reward}), where a decision step is executed when the agent receives a new workload to be offloaded. Hence, the simulation environment can progress and change by a number of simulation steps between two consecutive decision steps. This creates the concept of delayed rewards, where the reward $r_{n+1}$ of action $a_n$ can only be observed after a variable number of simulation steps. Because of the delayed nature of the reward in DES simulators, the observed reward is the immediate reward of the action that was performed in the previous decision step. This concept is essential in real deployments, which does not exist in simplified numerical experimentations since decision steps are simplified to simulation steps.

\begin{figure}[!t]
\centering
\includegraphics[trim=220 55 220 55, clip, width=0.4\textwidth]{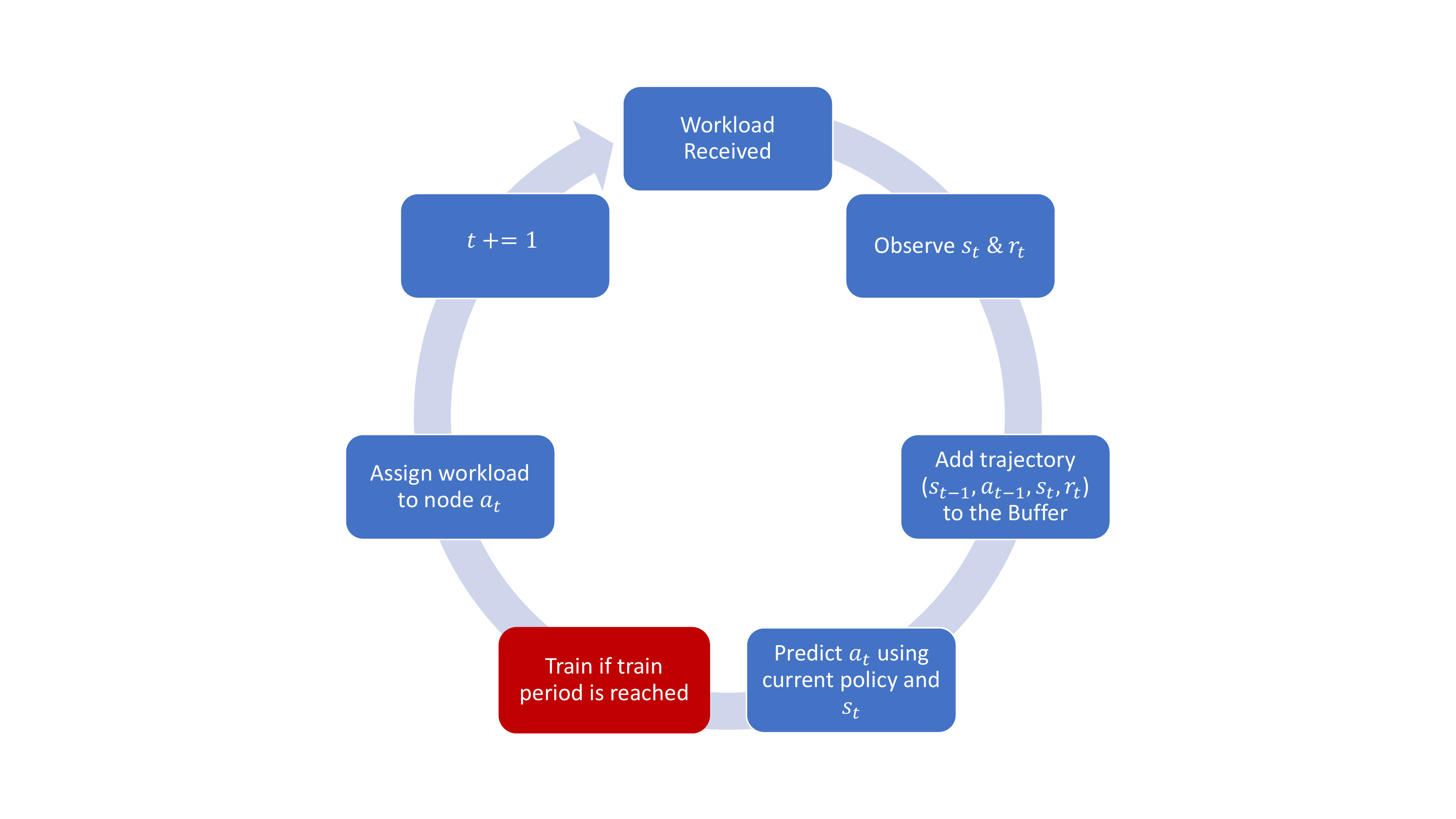}
\caption{A single Decision Step.}
\label{fig:decision}
\end{figure}

Figure \ref{fig:decision} shows what happens for our RL agent in a single decision step, which starts by receiving a workload and ends by assigning it to a Fog node according to the predicted action. First, the agent observes the current state and reward from the environment. To minimize the waiting delay, the immediate reward is defined by the increase/decrease in the number of jobs queued in all the nodes in the system (see Equation \ref{eq:reward}). Minimizing the time spent on the queues of computing nodes can significantly minimize the overall execution delay. 
\begin{equation}\label{eq:reward}
    r = \mathbb{Q}_{t-1} - \mathbb{Q}_{t} , \quad \text{where} \enspace \mathbb{Q} = \sum_{i=0}^N \mathtt{q}_i
\end{equation}

On the other hand, the state is defined in Equation \ref{eq:state} by the IoT cluster that generates the workload ($c$), the workload category ($w$), and the current distribution (normalized) of the load between Fog nodes ($d$). Discrete categories of workload requirements are used to simulate resource-demanding, moderate, and light workloads. While the normalized load distribution ($d$) represents the distribution of each workload category from every source cluster to every Fog node. Defining the state by not using resource and load information from Fog nodes provides privacy for Fog service providers. 
\begin{equation}\label{eq:state}
    s = <c, w, d> , \quad \text{where} \enspace \sum d = 1
\end{equation}

\begin{figure}[!t]
\centering
\includegraphics[trim=50 130 20 130, clip, width=0.48\textwidth]{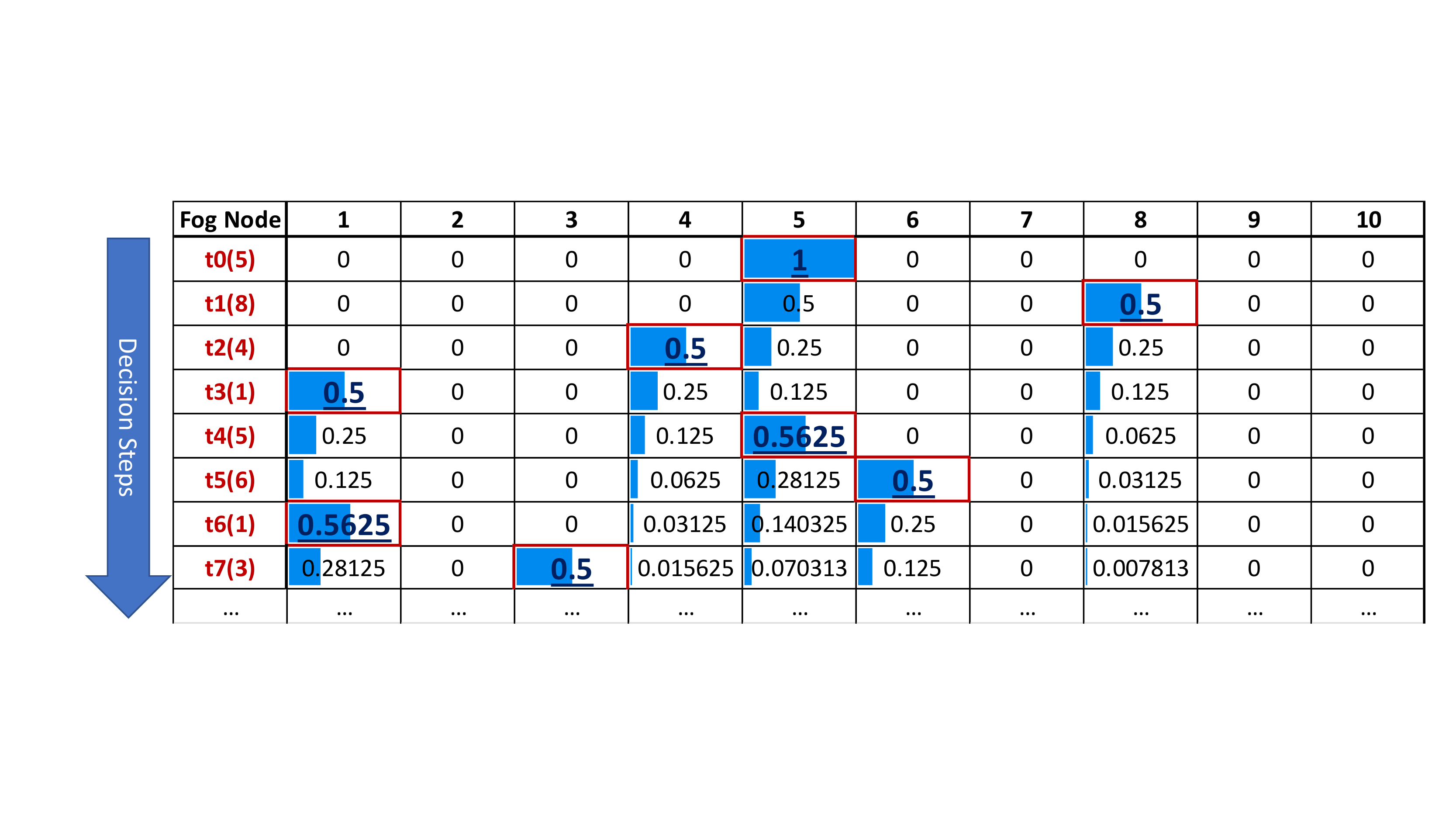}
\caption{Vanishing normalization for load distribution.}
\label{fig:vanishing}
\end{figure}

$d$ is implemented as a 3D matrix initialized to zeros with the following dimensions: number of actions ($|A|$) $\times$ number of clusters ($|C|$) $\times$ number of workload categories ($|W|$) (see Algorithm \ref{algo:LD}). Algorithm \ref{algo:LD} uses vanishing normalization, which gives higher values to most recent actions, and lower values for actions a larger number of steps away (see Fig. \ref{fig:vanishing}). This helps our agent see Fog nodes that have been selected recently for a given workload category from a given source cluster. It starts by adding 1 to the previous value of the array element indexed by action, cluster, and workload category of the previous decision step. Then, the array is normalized by dividing each element by the sum of the values of all array elements; after normalization, the sum of the values of all elements equals 1.

\begin{algorithm}
\caption{Load distribution with vanishing normalization.}
\label{algo:LD}
\begin{algorithmic}[1]
\renewcommand{\algorithmicrequire}{\textbf{Input:}}
\renewcommand{\algorithmicensure}{\textbf{Output:}}
\REQUIRE Previous action ($a$), source cluster ($c$), and workload category ($w$)
\ENSURE  Normalized load distribution ($d$)
\\ \textit{Initialisation} : \COMMENT{First decision step, no previous action}
\IF {(a = None)}
\STATE $d \gets \text{Zeros}(|A|, |C|, |W|)$
\RETURN $d$
\ENDIF
\\ \textit{Load Increment} :
\STATE $d(a, c, w) \gets d(a, c, w) + 1$
\\ \textit{Normalization} : \COMMENT{Divide each element by the total sum}
\STATE $d \gets \text{Normalize}(d)$
\RETURN $d$ 
\end{algorithmic} 
\end{algorithm}

\begin{figure}[!t]
\centering
\includegraphics[trim=140 140 90 60, clip, width=0.48\textwidth]{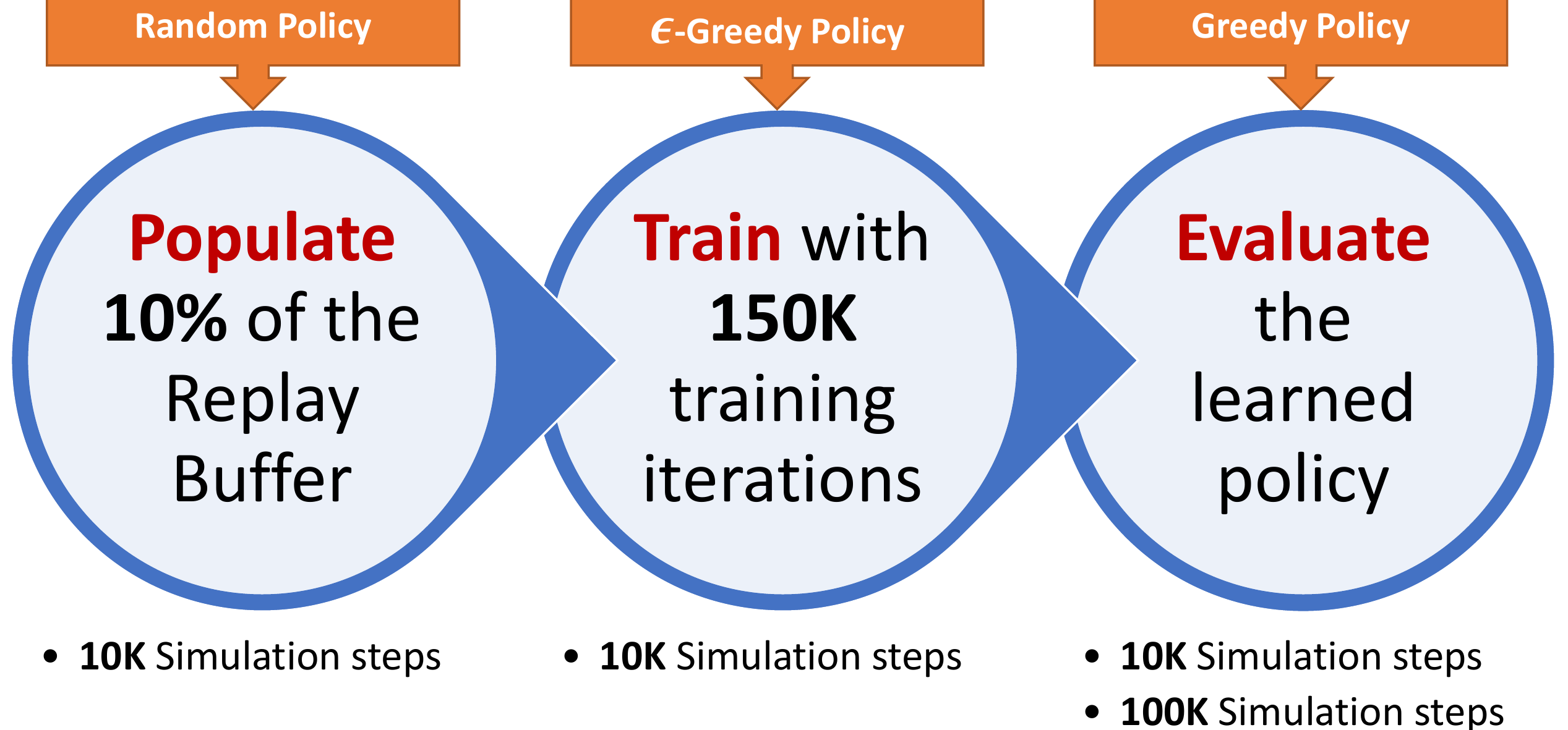}
\caption{Experimentation workflow.}
\label{fig:process}
\end{figure}

The second stage of the decision step in Figure \ref{fig:decision} is to add the trajectory (experience) to the replay buffer. A single trajectory is defined by the reward achieved after moving from the previous state to the next using the selected action. Using the current state of the environment, an action is then predicted using a random, Epsilon-Greedy, and greedy policy during buffer population, training, and evaluation of the agent (see Fig. \ref{fig:process}), respectively. During buffer population and training phases, episodes are created by ending the simulation after 10K simulation steps without a terminal state. On the other hand, the agent is evaluated with 10K and 100K simulation steps to show the agent ability to generalize in longer episodes. 

A decision step in the training phase includes training the neural network (marked in red in Fig. \ref{fig:decision}) that models the Action-Value ($Q$) of our agent. The neural network is trained with a random batch of trajectories from the replay buffer. This buffer is pre-populated in the buffer population phase and keeps growing in the training phase by the generated trajectories. The agent network is trained with 150K training steps, where a training step is performed after a predefined number of decision steps. The Network Train Period can be set to 1 decision step to train the agent after each decision step, but it can be also set to few more decision steps to allow a number of trajectories to be collected and added to the replay buffer before the next training step.

\section{Performance Evaluation}
\label{sec:evaluation}

\subsection{Simulation Parameters}
Using a Python-based DES Simulator, i.e., YAFS \cite{YAFS}, we evaluate our proposed approach on the realistic Fog environment used in our previous work \cite{ELECTRE}. First, a non-hierarchical graph is created from a simulated Autonomous System (AS) of the Internet \cite{networkx} to mimic flat Fog systems. Based on the betweenness centrality \cite{betweenness} of the generated nodes, each node is identified as either Cloud, Fog, or IoT source cluster. Similarly, Fog nodes are assigned compute resources inversely to their betweenness centrality to resemble unbalanced heterogeneous resources with unbalanced load distribution. We use to the common values of compute and network resources as in our previous work \cite{ELECTRE}, which are inspired from \cite{YAFS, Brogi2018DeployingFA, QoSIoT}.

As in our previous work \cite{ELECTRE}, three applications with different resource requirements simultaneously run in the system to simulate resource-demanding, moderate, and light workloads \cite{Brogi2021}. Each application workflow consist of two loops of interrelated services that interact to express the desired operational of realistic applications \cite{QoSIoT, Availability-Aware}. The first loop is an immediate Fog feedback for every generated workload (IoT$\rightarrow$Fog$\rightarrow$IoT). In the second loop, the Cloud aggregates 10\% of the workloads generated from the Fog nodes, and a feedback is sent back to the IoT source for 50\% of the aggregated workloads (IoT$\rightarrow$Fog$\xrightarrow{10\%}$Cloud$\xrightarrow{50\%}$IoT). Workloads are generated as a Poisson Point Process using exponential distributions with three scale parameters to simulate small, medium, and high workload generation rates, i.e., $\beta=200$, $\beta=150$, and $\beta=100$, respectively.

Our simulated Fog environment mimics the workflow of realistic IoT applications \cite{ELECTRE}, like online gaming, Internet of Vehicles (IoV), and health monitoring systems. We compare our proposed RL-based method against the ELECTRE method proposed in our previous study \cite{ELECTRE}. We also include four other baselines, namely random, Round-Robin, nearest node, and fastest service selection algorithms. The reader can consult the details in \cite{ELECTRE} for the implementation of these algorithms and the implementation of the simulated Fog environment. The proposed RL-based LB algorithm was implement using a DDQL agent tuned with the commonly used hyper-parameters shown in Table \ref{tab:paramsddqn}. The tuning of these hyper-parameters helped achieve the optimal performance of our agent in this particular setup.

\begin{table}[!t]
\caption{Tuned Hyper-parameters for the proposed DDQL agent \label{tab:paramsddqn}}
\centering
\begin{tabular}{|l||l|}
\hline
\textbf{Parameter} & \textbf{Values}\\
\hline
Discount Factor $\gamma$ & 0.99 \\
\hline
Decayed Epsilon-Greedy & The first 75\% Training steps \\
\hline
Decayed Epsilon values & Linearly from 100\% to 1\% \\
\hline
Replay Buffer Type & Uniform \\
\hline
Maximum Buffer Capacity & 1 Million trajectories of experience \\
\hline
Initial Buffer Population & 10\% of its maximum capacity \\
\hline
Buffer mini-Batch Size & 50 samples of 2 steps each \\
\hline
Network Train Period & 4 Decision steps \\
\hline
Target Update Period & 2000 Decision Steps \\
\hline
Network Layers & Fully connected [256, 128, 64] \\
\hline
Network Optimizer & Adam with $2.5e^{-4}$ learning rate \\
\hline
Network Loss Function & Huber Loss \\
\hline
\end{tabular}
\end{table}

\subsection{Performance Analysis}
\begin{figure}[!t]
    \centering
    \includegraphics[width=0.4\textwidth]{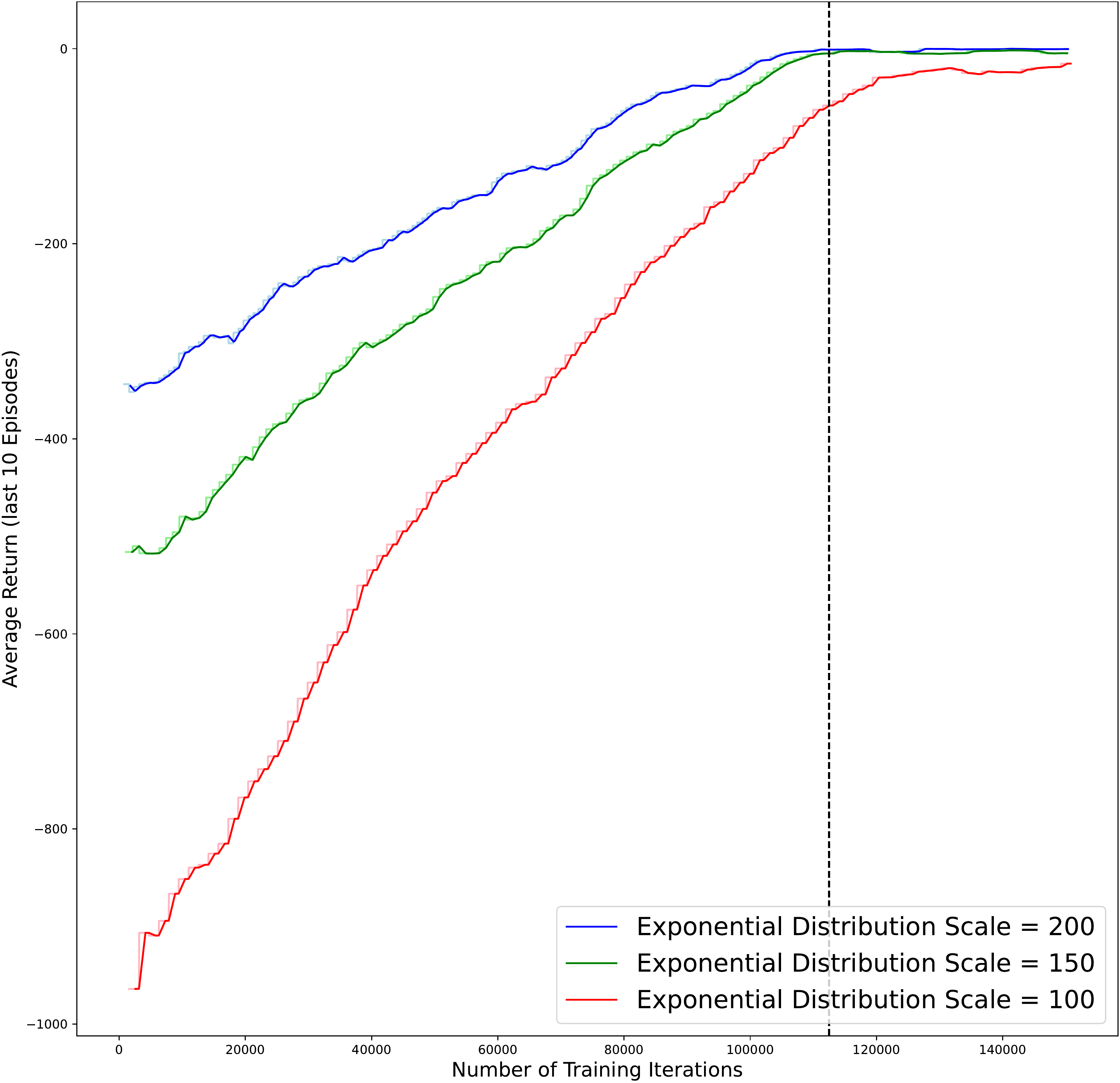}
    \caption{RL agent training performance.}
    \label{fig:avgR}
\end{figure}

Figure \ref{fig:avgR} shows the training performance, i.e., the average return in the last 10 training episodes, of our proposed RL agent. The performance gradually improves over the first 75\% of training iterations as indicated by the dotted black vertical line. This is due to the agent transformation from a fully randomized agent to an almost greedy agent with 1\% exploration probability, encouraging more exploration at the beginning of the learning process. The figure also shows that it is harder for the agent to learn when workloads have higher generation rates, i.e., exponential distribution with scale value of 100.

\begin{figure}[!t]
\centering
\subfloat[Exponential Distribution Scale = 200]{\includegraphics[width=0.48\textwidth]{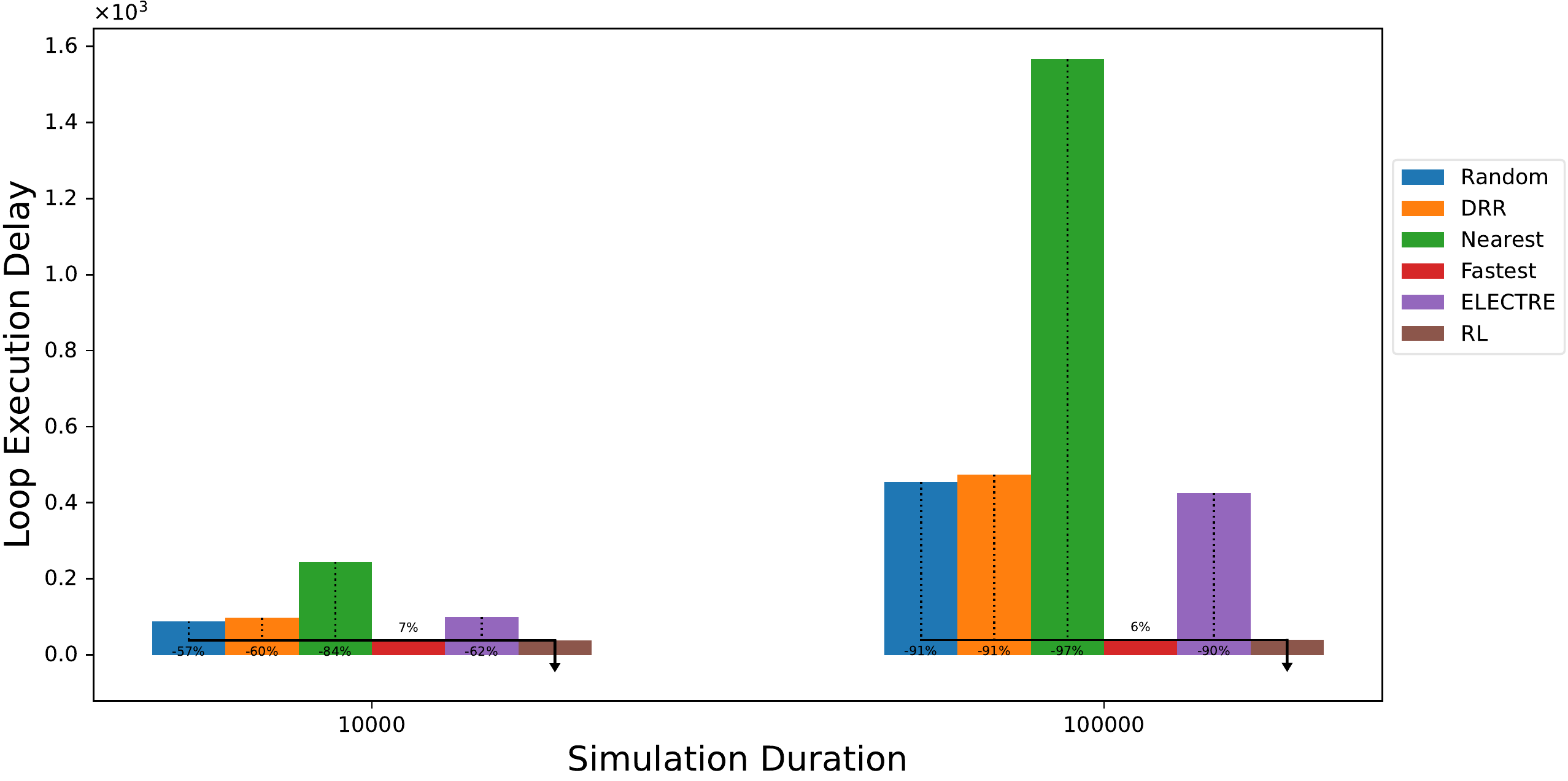}
\label{fig:lpEx200}}
\hfil
\subfloat[Exponential Distribution Scale = 150]{\includegraphics[width=0.48\textwidth]{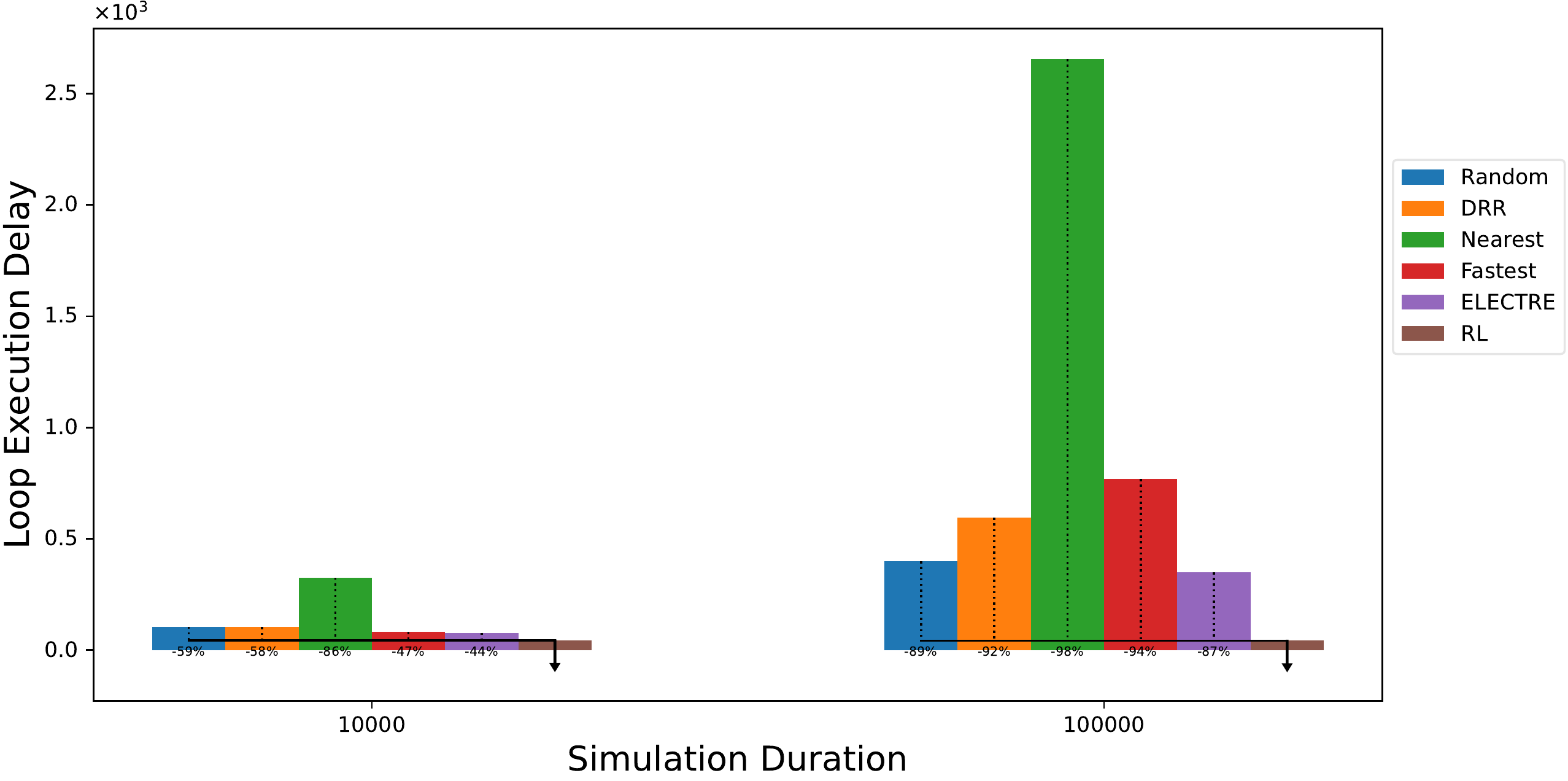}
\label{fig:lpEx150}}
\hfil
\subfloat[Exponential Distribution Scale = 100]{\includegraphics[width=0.48\textwidth]{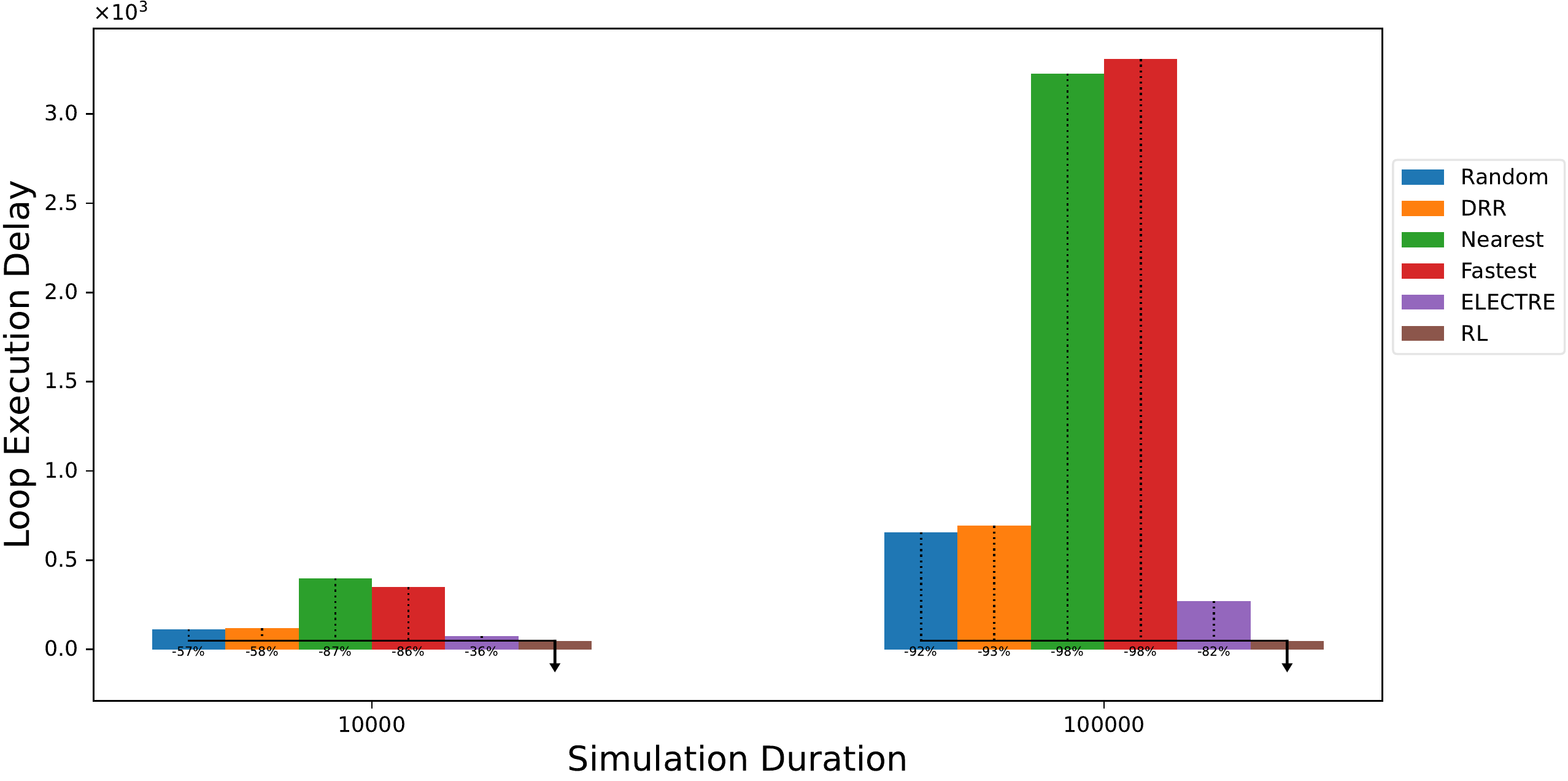}
\label{fig:lpEx100}}
\caption{Mean Loop Execution Delay.}
\label{fig:lpEx}
\end{figure}

Our proposed RL-based algorithm outperforms all other methods using every workload generation rate (see Fig. \ref{fig:lpEx}). Our DDQL agent improves the total execution delay over the ELECTRE algorithm by 82\% and 87\% using high and medium generation rates, respectively. Using the smallest generation rate, our method achieves a comparable performance with the fastest node selection algorithm (an improvement of 90\% over ELECTRE). This demonstrates that selecting the fastest node is good only with less frequent workloads. On the other hand, selecting the fastest node becomes the worst method with more frequent workloads. Figure \ref{fig:lpEx} also shows that our agent generalizes well when evaluated on $10\times$ more simulation steps. As explained in Section \ref{sec:method}, the agent is evaluated on 10K and 100K simulation steps, while it was only trained on 10K simulation steps.

\begin{figure*}[!t]
\centering
\subfloat[Workload Delays \cite{YAFS}]{\includegraphics[width=0.40\textwidth]{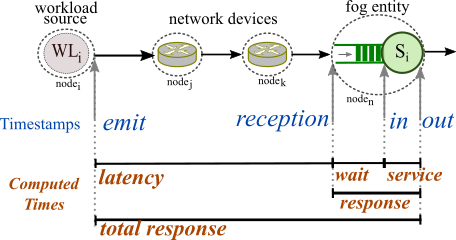}
\label{fig:yafs_times}}
\hfil
\subfloat[Latency]{\includegraphics[width=0.48\textwidth]{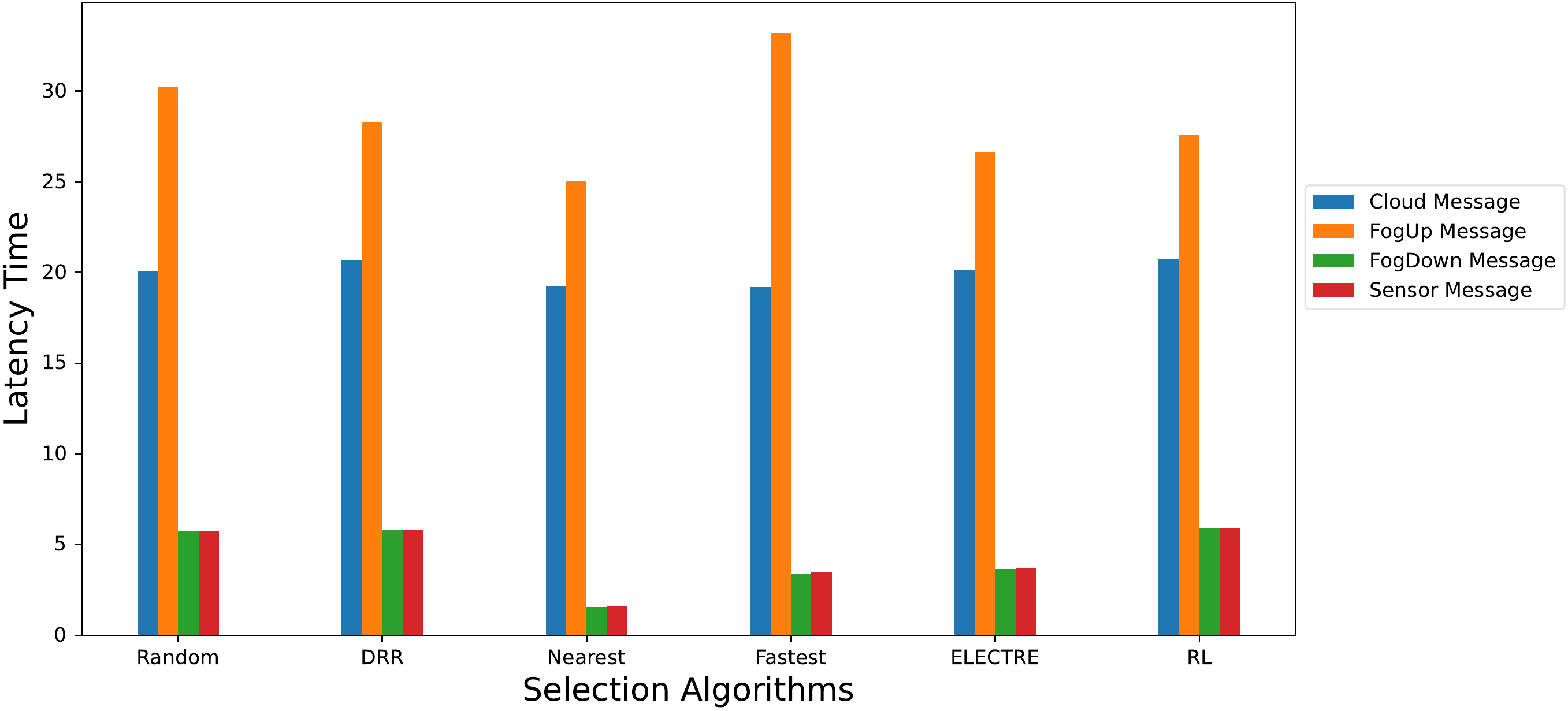}
\label{fig:time_latency}}
\hfil
\subfloat[Waiting Delay]{\includegraphics[width=0.48\textwidth]{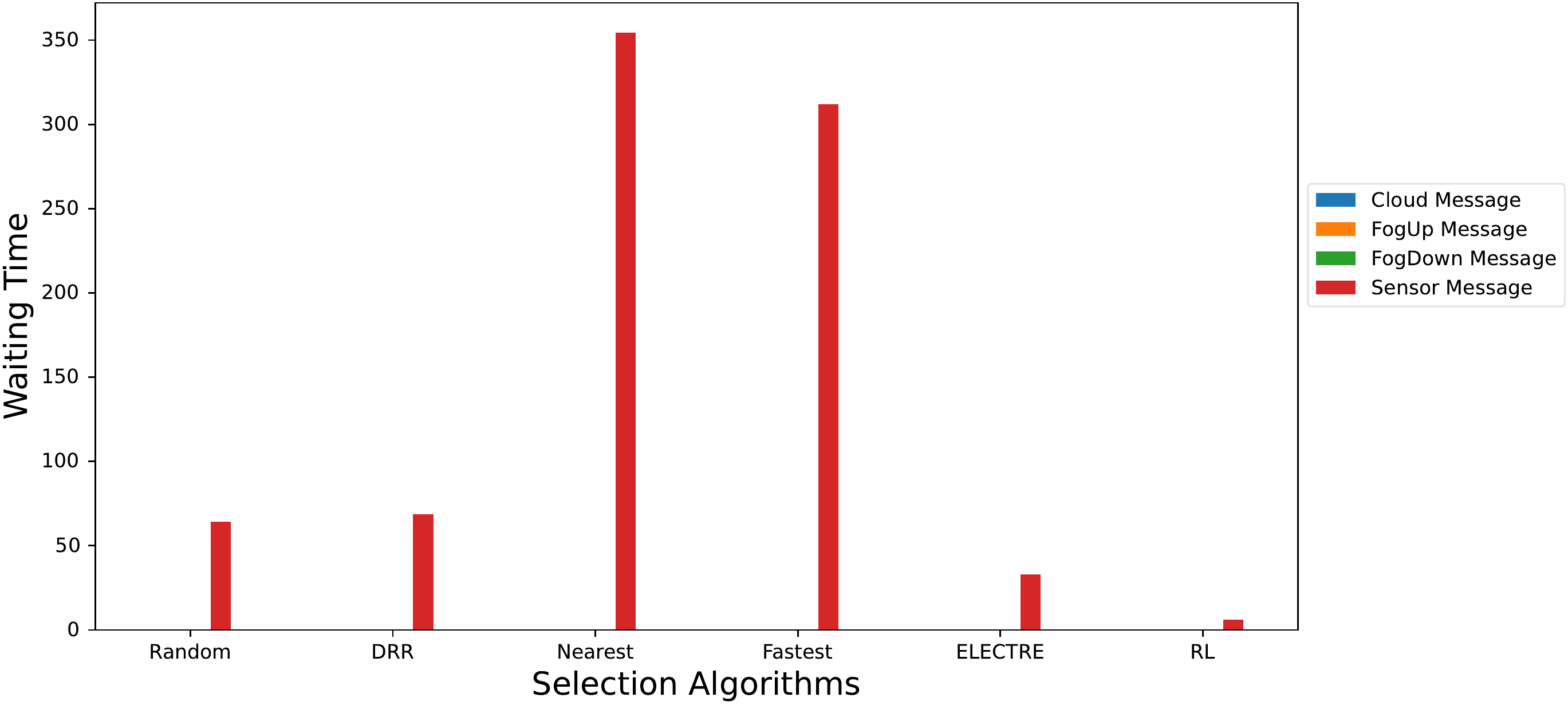}
\label{fig:time_waiting}}
\hfil
\subfloat[Service Time]{\includegraphics[width=0.48\textwidth]{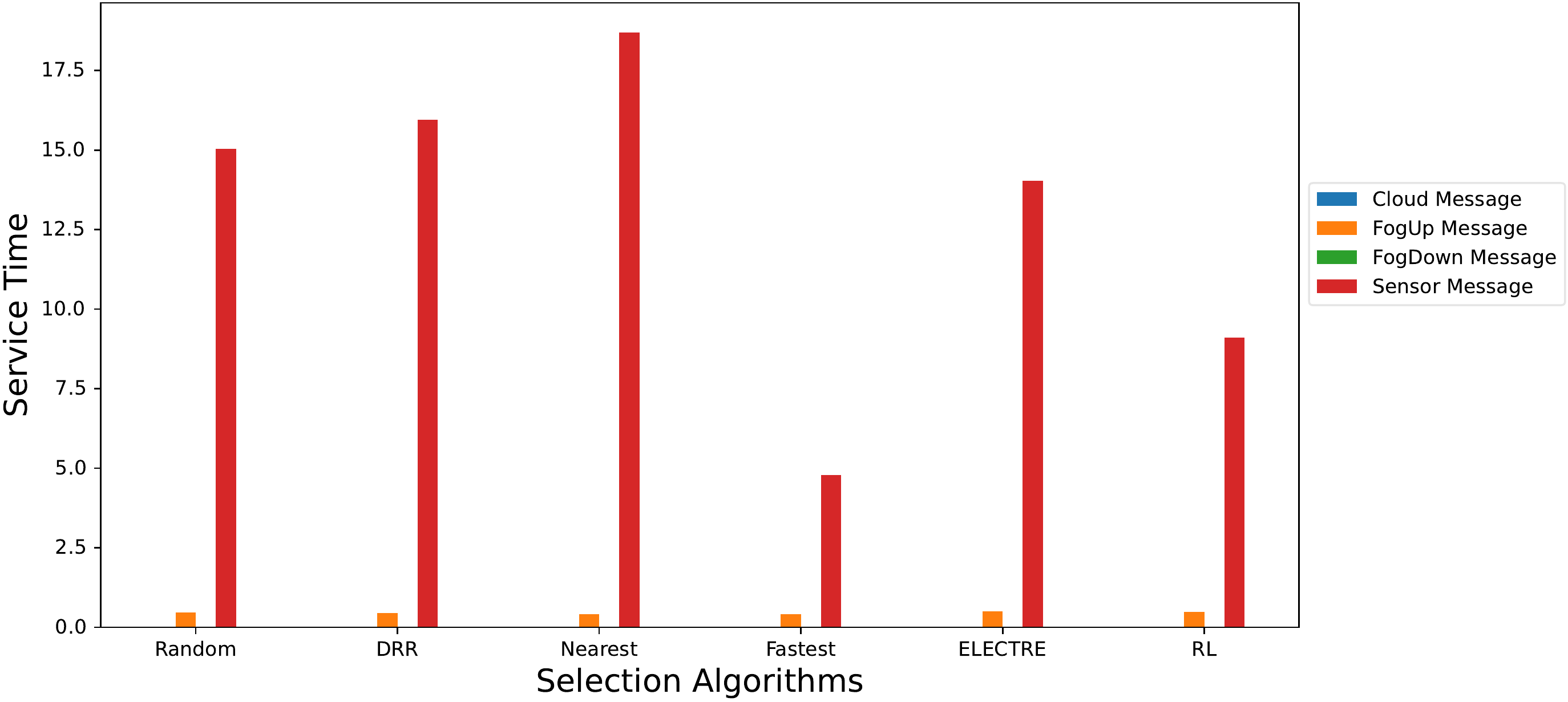}
\label{fig:time_service}}
\hfil
\subfloat[Response Time]{\includegraphics[width=0.48\textwidth]{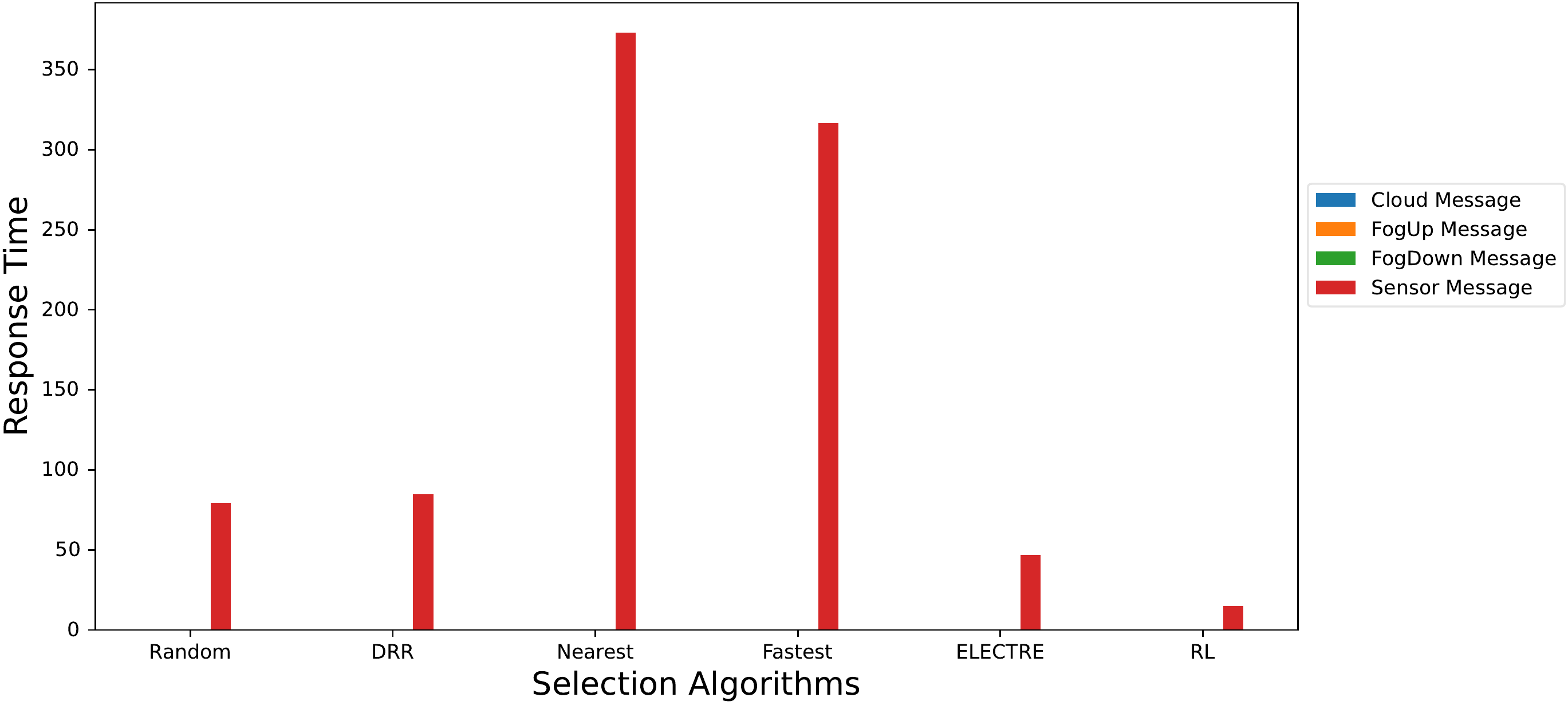}
\label{fig:time_response}}
\hfil
\subfloat[Total Response]{\includegraphics[width=0.48\textwidth]{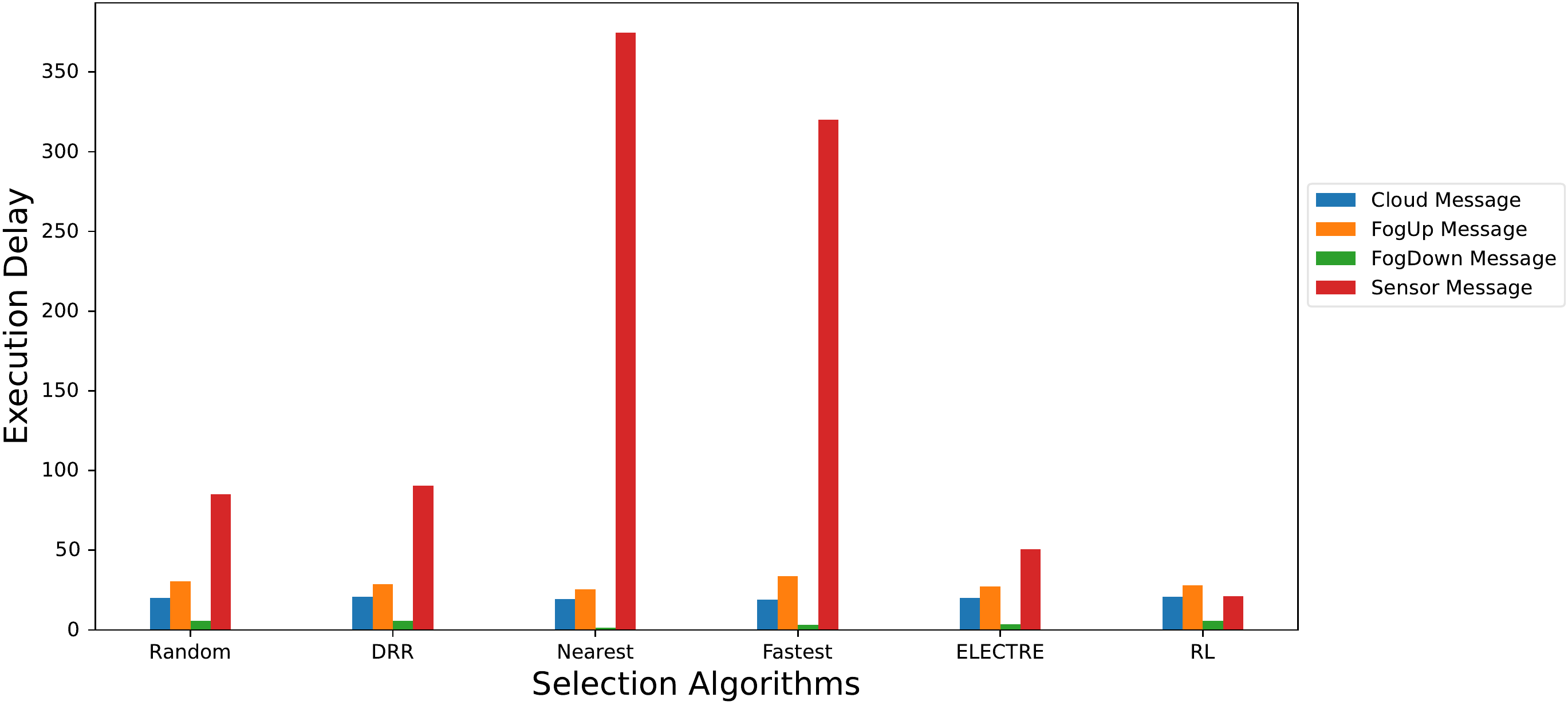}
\label{fig:time_total}}
\caption{Average workload delays.}
\label{fig:Delays}
\end{figure*}

To understand how our proposed method improves the overall system performance, we should first understand each of the five different delays shown in Fig. \ref{fig:Delays}\subref{fig:yafs_times} and their effect on the system. First, we have the network latency, which starts by emitting the workload from its source until it is received by its compute node. This depends on the network infrastructure, and hence, the fastest path between the source and destination is chosen for all selection algorithms.

Then, we have the response time, which is the time the compute node requires to process the workload, including waiting and processing times. The waiting time is the most critical factor for LB; a bad load distribution can cause the queue of some nodes to overflow while other nodes are ready to process more workloads. Hence, to improve the overall system performance, the proposed DDQL agent minimizes the number of workloads waiting in the queues of every computing node in the system simultaneously.

The proposed DDQL agent is able to minimize the waiting delay better than the other five methods used in this study (see Fig. \ref{fig:Delays}\subref{fig:time_waiting}). Figures \ref{fig:Delays}\subref{fig:time_latency} and \ref{fig:Delays}\subref{fig:time_service} show that the nearest node and fastest service selection algorithms, by nature, achieve the best latency and service delays, respectively. However, minimizing the accumulation of workloads in the queues of compute nodes is what enables our proposed RL-based method to achieve the best overall performance in terms of the response and the total response times (see Figures \ref{fig:Delays}\subref{fig:time_response} and \ref{fig:Delays}\subref{fig:time_total}), respectively).

\begin{figure*}[!t]
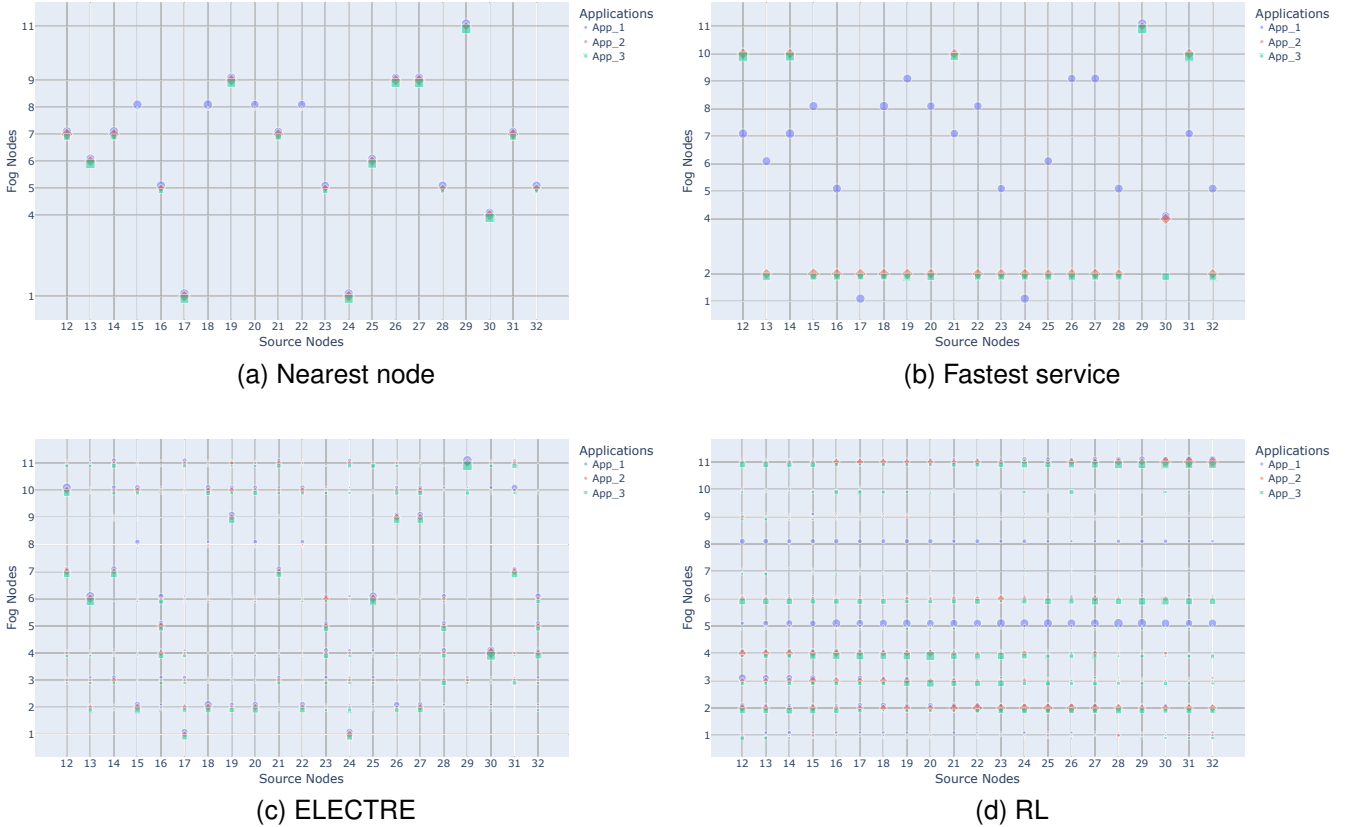

\centering
\subfloat[Nearest node]{\includegraphics[width=0.48\textwidth]{Figures/App_Distribution_Nearest_100.pdf}
\label{fig:NearestApps}}
\hfil
\subfloat[Fastest service]{\includegraphics[width=0.48\textwidth]{Figures/App_Distribution_Fastest_100.pdf}
\label{fig:FastestApps}}
\hfil
\subfloat[ELECTRE]{\includegraphics[width=0.48\textwidth]{Figures/App_Distribution_ELECTRE_100.pdf}
\label{fig:ELECTREApps}}
\hfil
\subfloat[RL]{\includegraphics[width=0.48\textwidth]{Figures/App_Distribution_RL_100.pdf}
\label{fig:RLApps}}
\caption{The distribution of workloads among the Fog nodes.}
\label{fig:AppDist}
\end{figure*}

Figure \ref{fig:AppDist} shows the distribution of workloads from IoT source clusters to every Fog node. The nearest node selection algorithm assigns all workloads from an IoT source cluster to the Fog node that is directly connected to it (see Fig. \ref{fig:AppDist}\subref{fig:NearestApps}). On the other hand, the fastest service selection algorithm calculates the total execution delay for the generated workload based on the workload requirements and the resource capabilities of Fog nodes. Then, it assigns the workload to the Fog node that has the smallest total execution delay for that workload (see Fig. \ref{fig:AppDist}\subref{fig:FastestApps}).

For light workloads, i.e., $App_1$, the fastest service is always in the Fog node directly connected to the IoT source cluster. That is because Fog nodes with higher resources, i.e., $Fog_2$ \& $Fog_{10}$, are almost always selected for resource-demanding workloads using the fastest service selection algorithm. Unlike these basic load assignment mechanisms, the ELECTRE algorithm achieves a good performance by distributing workloads from each Iot source cluster between multiple Fog nodes to increase their resource utilization \cite{ELECTRE} (see Fig. \ref{fig:AppDist}\subref{fig:ELECTREApps}). 

This behavior minimizes the number of waiting requests in each Fog node, which allows ELECTRE to achieve a better waiting delay compared to the other baselines. However, a better distribution of workloads between Fog nodes is achieved using our proposed RL-based method (see Fig. \ref{fig:AppDist}\subref{fig:RLApps}), which tends to lower the number of waiting tasks in the queues of compute nodes even further. Minimizing the queue length with such optimal workload distribution allows our RL-based method to lower the waiting delay in all compute nodes. This lowers the total execution delay of each application loop to improve the overall system performance.

To demonstrate the impact of the queue length in minimizing the total execution delay, we compare our privacy-aware RL solution (PARL) against privacy-lacking RL solutions (PLRL) from the literature \cite{Managing, ReTra}. In \cite{Managing, ReTra}, the state is represented by the node that needs to allocate a task to a Fog node and the number of tasks currently remaining in the queue of each Fog node; i.e., queue length (current load). \cite{Managing, ReTra} aim to minimize the processing delay and overload probability of Fog nodes. Hence, their reward function is a negated sum of the immediate execution delay and an overflow indicator using the current load of the selected Fog node.

We implemented the state representation of \cite{Managing, ReTra}, which requires the current load information of all Fog nodes in the system. Besides the reward function in \cite{Managing, ReTra}, we also implemented the reward using only the Execution Delay or the Queue Length. This will show what contributes the most to minimize the total loop execution delay. Hence, we evaluated the following three flavors for the reward function, where ED and QL are calculated using resource and load information, respectively.
\begin{itemize}
  \item The Execution Delay (ED) of the recently assigned task.
  \item The Queue Length (QL) in the recently chosen Fog node.
  \item The sum of Execution Delay \& Queue Length (ED+QL).
\end{itemize}

\begin{figure*}[!t]
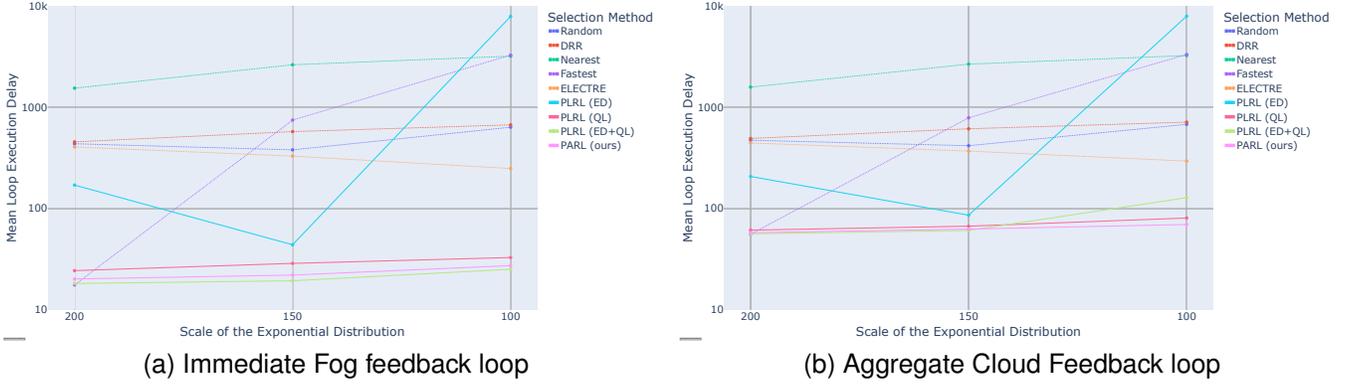

\centering
\subfloat[Immediate Fog feedback loop]{\includegraphics[width=0.48\textwidth]{Figures/Fog_Loop.pdf}
\label{fig:FogLoop}}
\hfil
\subfloat[Aggregate Cloud Feedback loop]{\includegraphics[width=0.48\textwidth]{Figures/Cloud_Loop.pdf}
\label{fig:CloudLoop}}
\caption{Comparing the performance of traditional approaches against various state/reward representations for RL load balancing algorithms.}
\label{fig:CompareRL}
\end{figure*}

Figure \ref{fig:CompareRL} shows that PARL and PLRL achieve better results than traditional load distribution schemes. The only exception is PLRL (using ED in the reward function), where the immediate reward of the most recent action is the execution delay of the recently assigned task. This is because when the execution delay is the reward, the RL agent learns to select the node with the smallest execution delay, i.e., the fastest Fog node. With higher workload generation rates, more and more workloads will be congested on these fast nodes while wasting resources in slower nodes. 

Figures \ref{fig:CompareRL}\subref{fig:FogLoop} and \ref{fig:CompareRL}\subref{fig:CloudLoop} show the algorithm performance on the two distributed application loops, i.e., Immediate Fog feedback loop and Aggregate Cloud Feedback loop, respectively. Traditional load distribution schemes are identified in the figures with dashed lines while RL-based algorithms are identified with solid lines. The $x$-axis represents the three scale parameters that simulate small ($\beta=200$), medium ($\beta=150$), and high ($\beta=100$) workload generation rates.

Figure \ref{fig:CompareRL} shows that the fastest node selection algorithm achieves the best performance with the smallest workload generation rate. However, its performance gets worst with higher workload generation rates. On the other hand, RL-based algorithms consistently show a better performance even with higher workload generation rates. The only exception is when the RL agent selects the node with the smallest execution delay. In this case, the RL agent tries to select the fastest Fog node for the task. However, it performs worst than the simple fastest node selection method because of the randomness caused by unseen states that are approximated by the neural network. Hence, minimizing the number of queued tasks in Fog nodes, i.e., queue lengths, is the major factor to minimize the total loop execution delay. 

Using the execution delay of the recently assigned task and queue length of the recently chosen Fog node together (ED+QL) helps slightly improve the performance of PLRL algorithms (see Fig. \ref{fig:CompareRL}). The performance of our proposed PARL state/reward representation is on bar with the results of the PLRL (ED+QL) solution. However, our proposed representation has an additional benefit of using neither load nor resource information from Fog nodes, which helps maintain the privacy required by Fog service providers. In addition, with the highest workload generation rate, our proposed PARL solution outperforms all other methods on the Cloud application loop. This is because it learns to minimize the number of queued tasks in all compute nodes in the system, including the Cloud itself.

\section{Conclusion}
\label{sec:conclusion}

% In this article, a privacy-aware load balancing algorithm was proposed to improve the overall system performance by minimizing the number of waiting tasks in the queues of all compute nodes in the system. To maintain privacy, the proposed algorithm does not require load and resource information from Fog nodes, which helps the algorithm to dynamically adapt to possible changes in the environment. To do this, our proposed DDQN agent uses the change in the number of waiting jobs between two consecutive decision steps as the immediate reward of its most recent action. In addition, it represents its state as a normalized distribution of workload categories from source clusters to Fog nodes, which vanishes over time to emphasize recent decisions.

% To mimic the practical deployment of our solution in real environments, we interactively evaluate our agent using a DES simulator with more simulation steps than what it was initially trained with. The results of our experiments showed that our proposed DDQN algorithm outperforms traditional baselines used in the literature as well as a search-based optimization method, i.e., ELECTRE. This work shows the ability of RL methods to learn in partially observable environments, where other environment information is not accessible or is difficult to be modeled. In future work, we will evaluate the benefits of transfer learning techniques for RL methods, which can help reduce the time required to retrain the agent in case of possible changes in the environment.

In this paper, we propose a load balancing algorithm that is privacy-aware to improve overall system performance by reducing the number of waiting tasks in compute nodes' queues. The proposed algorithm does not compromise privacy since it does not require load and resource information from Fog nodes. Specifically, the proposed DDQN agent uses the change in the number of waiting jobs as the immediate reward of its most recent action; it represents its state using a normalized distribution of workload categories, which vanishes over time to emphasize recent decisions. The interactive evaluation of our agent using a DES simulator, with more simulation steps than what it was initially trained with, mimics practical deployment in real environments. 

Our experiments demonstrate that our proposed DDQN algorithm outperforms traditional baselines used in the literature as well as a search-based optimization method, i.e., ELECTRE. Our results also indicate that our proposed privacy-aware environment representation for the RL agent outperforms other representations from the literature that compromise privacy. This work highlights the ability of RL methods to learn in partially observable environments, where other environment information is not accessible or is difficult to be modeled. In future work, we plan to evaluate transfer learning techniques for RL methods, which can help reduce the time required to retrain the agent in case of possible changes in the environment, while ensuring privacy.

\section*{Acknowledgments}
The authors would like to express their sincere gratitude to Dr. Pierre-Luc Bacon for his valuable guidance and support to this research process. Dr. Bacon expertise in Reinforcement Learning greatly contributed to the success of this project.

\bibliographystyle{IEEEtran}
\bibliography{main.bib}

\begin{IEEEbiography}[{\includegraphics[width=1in,height=1.25in,clip,keepaspectratio]{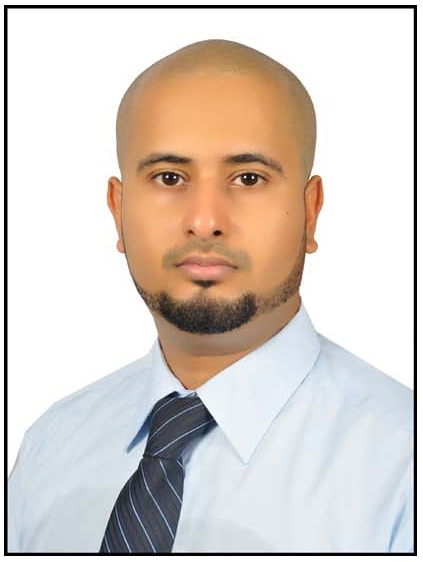}}]{Maad Ebrahim} is currently a Ph.D. candidate at the Department of Computer Science and Operations Research (DIRO), University of Montreal, Canada. He received his M.Sc. degree in 2019 from the Computer Science Department, Faculty of Computer and Information Technology, Jordan University of Science and Technology, Jordan. His B.Sc. degree in Computer Science and Engineering has been received from the University of Aden, Yemen, in 2013. His research experience includes Computer Vision, Artificial Intelligence, Machine learning, Deep Learning, Data Mining, and Data Analysis. His current research interests include Fog and Edge Computing technologies, Internet of Things, Reinforcement Learning, and Blockchains.
\end{IEEEbiography}

\begin{IEEEbiography}[{\includegraphics[width=1in,height=1.25in,clip,keepaspectratio]{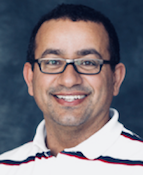}}]{Abdelhakim Hafid}
spent several years as the Senior Research Scientist with Bell Communications Research (Bellcore), NJ, USA, working in the context of major research projects on the management of next generation networks. He was also an Assistant Professor with Western University (WU), Canada, the Research Director of Advance Communication Engineering Center (venture established by WU, Bell Canada, and Bay Networks), Canada, a Researcher with CRIM, Canada, the Visiting Scientist with GMD-Fokus, Germany, and a Visiting Professor with the University of Evry, France. He is currently a Full Professor with the University of Montreal. He is also the Founding Director of the Network Research Laboratory and Montreal Blockchain Laboratory. He is a Research Fellow with CIRRELT, Montreal, Canada. He has extensive academic and industrial research experience in the area of the management and design of next generation networks. His current research interests include the IoT, Fog/Edge Computing, blockchain, and intelligent transport systems.
\end{IEEEbiography}

\vfill
\end{document}